\documentstyle[12pt]{article}
\newcommand{\newsection}{
\setcounter{equation}{0}
\section}
\newcommand{\rf}[1]{(\ref{#1})}
\newcommand{\bea}{\begin{eqnarray}}
\newcommand{\eea}{\end{eqnarray}}
\newcommand{\g}{\gamma}

\renewcommand{\b}{\beta}
\renewcommand{\a}{\alpha}

\newcommand{\m}{\mu}

\newcommand{\sg}{\sigma}

\def\void{}
\def\labelmark{}

\newenvironment{formula}[1]{\def\labelname{#1}
\ifx\void\labelname\def\junk{\begin{displaymath}}
\else\def\junk{\begin{equation}\label{\labelname}}\fi\junk}%
{\ifx\void\labelname\def\junk{\end{displaymath}}
\else\def\junk{\end{equation}}\fi\junk\labelmark\def\labelname{}}

{\ifx\void\labelname\def\junk{\end{array}\end{displaymath}}
\else\def\junk{\end{array}\right.\end{equation}}
\fi\junk\labelmark\def\labelname{}\def\junk{}
}

\newcommand{\beq}{\begin{formula}}
\newcommand{\eeq}{\end{formula}}
\newcommand{\beqv}{\begin{formula}{}}

\begin{document}
\topmargin 0pt
\oddsidemargin 5mm
\headheight 0pt
\headsep 0pt
\topskip 9mm

\hfill    NBI-HE-93-4

\hfill February 1993
\begin{center}

\vspace{24pt}

{\large \bf

Baby Universes in 2d Quantum Gravity}

\vspace{24pt}

{\sl Jan Ambj\o rn}

\vspace{6pt}

 The Niels Bohr Institute\\
Blegdamsvej 17, DK-2100 Copenhagen \O , Denmark\\

\vspace{12pt}

{\sl Sanjay Jain}

\vspace{6pt}

 Centre for Theoretical Studies \\
 Indian Institute of Science \\
 Bangalore 560012 , India\\

\vspace{12pt}

{\sl Gudmar Thorleifsson}

\vspace{6pt}

 The Niels Bohr Institute\\
Blegdamsvej 17, DK-2100 Copenhagen \O , Denmark\\

\end{center}

\addtolength{\baselineskip}{0.20\baselineskip}

\vspace{12pt}

\vfill

\begin{center}
{\bf Abstract}
\end{center}

\vspace{6pt}

\noindent
We investigate the fractal structure of $2d$
quantum gravity, both for pure gravity and for gravity coupled to
multiple gaussian fields and for gravity coupled to Ising spins.
The roughness of the
surfaces is described in terms of baby universes and using numerical
simulations we measure their distribution which is related to
the string susceptibility exponent $\g_{string}$.

\vfill

\newpage

\newsection{Introduction}

The fractal and selfsimilar structure of $2d$ quantum gravity is
related to the entropy- (or string susceptibility-) exponent $\g$.
This has been discussed in a recent paper \cite{jain} where the
structure of so-called baby universes was analyzed. It is convenient
in the following discussion to consider $2d$ quantum gravity with an
ultraviolet cut-off and we will
consider the surfaces entering in the path integral as triangulated
surfaces built out of equilateral triangles \cite{david,kkm,adf}. In
the case of surfaces of spherical topology a closed, non-intersecting
loop along the links will separate the surface in two parts. The smallest
such loop will be of length 3. It will split the surface in two parts. If
the smallest part is different from a single triangle we call it
(following the notation of \cite{jain}) a ``minimum neck baby universe'',
abbreviated ``mimbu''. The smallest possible area of a mimbu is 3 and
the largest possible area will be $N_T/2$, where $N_T$ is the number of
triangles constituting the surface.

In the case of pure 2d quantum gravity it is known that the number of distinct
surfaces of genus zero made out of $N_T$ triangles
has the following asymptotic form
\beq{*1.1}
Z(N_{T}) \sim e^{\mu_c N_{T}} N_{T}^{\g-3}
\eeq
where $\g= -1/2$. For the models which can be solved explicitly
and where  $c<1$ we have the following partition function:
\beq{*1.1a}
Z(\m) = \sum_{N_T} Z(N_T) e^{-\m N_T}
\eeq
where $Z(N_T)$ for large $N_T$ is of the form \rf{*1.1}, just with a different
$\g= \g(c)$. For $c=1$ it is known that there are logarithmical corrections
to \rf{*1.1}, while the asymptotic form of $Z(N_T)$ is unknown for
$c >1$, although it can be proven
that it is exponentially bounded (\cite{adf}).
If we {\it assume} \rf{*1.1}  one can prove that the average number
of mimbu's of area $B$ on a closed surface of spherical topology and
with area $N_T$ (we use the notation $area \equiv \# triangles$) is given by
\beq{*1.2}
\bar{n}_{N_{T}}(B) \sim (N_{T} - B)^{\g-2} B^{\g-2}
\eeq
provided $N_T$ and $B$ are large enough.

\vspace{12pt}

\noindent
The above formalism is well suited for numerical simulations. The measurement
of the exponent $\g$ has always been somewhat difficult. The first attempts
used a grand canonical updating (\cite{adfo,jkp,afkp}), which generated
directly the distribution  \rf{*1.1a}. The disadvantage is that one has
to fine-tune the value of $\m$ to $\m_c$. Later improved versions allowed
one to avoid this \cite{abk}, but one still
had to perform independent Monte Carlo
simulations for a whole range of $N_T$ and $\g$ still appeared as a
subleading correction to the determination of the critical point $\m_c$.
These disadvantages disappear when we use \rf{*1.2}. $\g$ does not appear
as a subleading correction to $\m_c$ and one can use canonical Monte Carlo
simulations (the so-called link-flip algorithm  \cite{kkm}) which
keeps $N_T$ fixed, and still in a single Monte Carlo
simulation get a measurement of the distribution of mimbu's all the way
up to $N_T/2$. One can therefore take a large $N_T$ and make one very long
run. This allows us to avoid the problems with long thermalization time.
In addition the actual measurement of the mimbu distribution is easy.
For a given thermalized configuration one has to identify all possible
mimbu's associated with the configuration. This is done by
picking up one link,
$l_0$, and checking whether any  links which have a vertex in common with
$l_0$ have a vertex in common which do not belong to $l_0$.
This being the case
we will have a minimal neck of length 3. For a given $l_0$ there will always
be two such, corresponding to the two triangles sharing $l_0$. But there
might be additional ones and they will divide the surface into a mimbu
and its ``mother''. By scanning over $l_0$'s, avoiding double counting and
repeating the process for independent configurations we can construct
the distribution of mimbu's.

\vspace{12pt}

\noindent
In the rest of this paper we report on the results of such
numerical simulations.

\newsection{Numerical simulations}

\subsection{Pure gravity}

The simulations were done on lattices of size ranging from 1000 to 4000
triangles $N_{T}$ and we used the standard ``link flip'' algorithm
\cite{kkm} to update the geometry.
We used of the order of
$10^{7}$ sweeps, where each sweep consists of $N_{T}$ link flips.
After thermalization we measured for each $10^{th}$  sweep the distribution
of mimbu's, that is we counted all areas
$B>1$ enclosed by boundaries of length 3.  The reason for performing the
measurements so often is simply that they are not time consuming (the time
it takes to make one measurement
is comparable to the time it takes to perform  one sweep).
The distributions are shown in fig. 1.
In order to extract $\g$ the distributions are fitted
to equation \rf{*1.2}.
But as eq. \rf{*1.2} is only asymptotically correct
deviations can be expected for
small $B$. Thus a lower cut-off $B_{0}$ has to be introduced in
the data to avoid the effects of this deviations.
Moreover  we have added the simplest type of correction term
which arises from the replacement
\beq{*2.0}
B^{\g-2} \to B^{\g-2}\left( 1+ \frac{C}{B} + {\cal O} (1/B^2) \right)
\eeq
in \rf{*1.2}  and fitted to the form
\beq{*2.1}
\ln (\bar{n}_{N_{T}}) = A + (\g-2) \ln(B(1-\frac{B}{N_{T}}))
+ \frac{C}{B}
\eeq
for $B \geq B_{0}$.  $A$ and $C$ are some fit parameters.
Comparison of the results with and without this
correction term can be seen
in fig. 2 where we plot the value of $\g$ extracted with
different cut-off's $B_{0}$.
We see that including the correction improves the results
considerable.

Let us  assume that the values $\g_{B_0}$ extracted from \rf{*2.1} appproach
exponentially a limiting value for large $B_0$:
\beq{*ex1}
\g_{B_0} = \g -c_1 e^{-c_2 B_0}.
\eeq
The result of such a fit is shown in fig. 2. It is clear from fig. 2
that the assumption of an exponential approach of $\g_{B_0}$ to $\g$
is not essential for the extraction of $\g$. We have introduced it at this
point in order to treat all measurements consistently. For the matter
fields coupled to gravity the finite size effects will be larger and
extrapolation to large $B_0$ more important.

The $\g$ extracted in this way for different lattice sizes is:

\vspace{8pt}
\begin{center}
\begin{tabular}{cc}
$N_{T}$  &  $\g$     \\ \hline
$1000$   &  $-0.496 \pm 0.005$  \\
$2000$   &  $-0.501 \pm 0.004$  \\
$3000$   &  $-0.504 \pm 0.004$
\end{tabular}
\end{center}
\vspace{8pt}

\noindent
which is in good agreement with the expected value of $\g = -0.5$.
It shows that this kind of simulations are indeed well suited
to measure $\g$ and it is thus natural to try apply them to the
case of matter couple to $2d$ gravity.

\subsection{The Ising model}

The next non-trivial test of the method is to study the Ising model
coupled to $2d$ gravity.
It has been solved analytically \cite{kazakov} and was found
to have a $3rd$-order phase transition.
The coupling to gravity is in a sense weak as it
only changes the string susceptibility {\it at} the
critical point (from $\g = -1/2$ to $\g = -1/3$).
For this reason it has until now been considered very difficult
to measure $\g$ directly, since it required a fine-tuning of both
the bare cosmological constant $\m$ and the spin coupling constant $\b$.
On the other hand it has been verified that it is
indeed possible to extract the other known critical exponents \cite{ising}
since  for these exponents it is possible to use
the canonical ensemble in the simulations.

The Ising spins are placed in the center of the triangles
and they  interact with the spins on neighbouring triangles.
This corresponds to  placing them on vertices in the dual graph. In
that case the critical point has been found explicitly
and is $\b_{c} = 0.7733...$ \cite{bj}.
The (canonical) partition function of the model is
\beq{*2.3}
Z_{N_{T}}(\b) = \sum_{T\sim N_T} \sum_{\{\sigma_{i}\}}
e^{\b \sum_{<i,j>} \sg_{i}\sg_{j}}
\eeq
where the summation is over all triangulations with $N_T$ triangles.

In the simulations we used a Swendsen-Wang cluster algorithm \cite{SW}
to update the Ising spins and lattices sizes
$N_{T} = 1000$ and $2000$.
We made runs for several values of
the coupling in the interval $0.6 \leq \b \leq 0.95$ and
then fitted the distributions to eq. \rf{*2.1}. In this
way we could extract values $\g_{B_0}(\b)$ and by assuming a relation
like \rf{*ex1}:
\beq{*ex2}
\g_{B_0} (\b) = \g(\b) -c_1(\b) e^{-c_2(\b) B_0}
\eeq
we have extracted the values for $\g(\b)$ shown in fig. 3. Examples of
$\g_{B_0}(\b)$ and the exponential fit \rf{*ex2} for different values of
$\b$ are shown in fig. 4. We observe a marked increase in the
dependence on $B_0$ when $\b$ approaches $\b_c$.

We get, as expected, the pure gravity value of $-0.5$ for
couplings far below and above $\b_{c}$.  In the vicinity
of the phase transition we see on the other
hand a clear peak and the peak values agree well with
the exact value $\g = -1/3$. We conclude that the method for extracting
$\g$ works well in this case too, although it should be clear that the
amount of numerical work needed is much larger in this case than in
the case of pure gravity.

\subsection{Gaussian fields}

The gaussian fields $x^\m$, $\m=1,\dots,D$ are placed
on the sites $i$ of the triangulation $T$. They
can be viewed as representing an immersion $i \to x_i^\m$
of our abstract triangulation $T$ into $R^D$, i.e. a model for
non-critical strings and they also represent a coupling of matter
with central charge $c=D$ to gravity.
The multiple gaussian models
do not interact directly with each other but
only through their mutual interaction with the geometry.
The (canonical) partition function is given by
\beq{*2.2}
Z_{N_{T}} = \sum_{T \sim N_T}
\int \prod_{i{\in}T\backslash \{ i_{0} \}} d^D x_{i}
\: e^{ -\sum_{<i,j>} (x_{i}^\m - x_{j}^\m)^2}
\eeq
where the summation is over all triangulations $T$ with $N_T$ triangles.
One site is kept fixed in order to eliminate the translation mode.
No coupling constant appears in the action as it can be absorbed
in a redefinition of the gaussian fields.

Again we have performed simulations with up to $10^{7}$ sweeps for
lattice sizes ranging from 1000 to 4000 triangles.
We have used from one to five gaussian fields
and a standard  Metropolis algorithm to update them. In fig. 5
we show how the distributions of baby universes change with
increased $c$ (normalized with the distribution for pure gravity).
Fitting these distributions to the functional form  \rf{*2.1}
and extracting $\g$ as above yields the results shown in fig. 6.
The results are compatible with earlier estimates \cite{abk}.

\vspace{12pt}

\noindent
It is seen that $\g$ is too small for $c=1$ where it is known that $\g=0$.
But in the case $c=1$ we know that the asymptotic form
\rf{*1.1} is not correct. It should be multiplied with logarithmic
corrections. If we include these we get for $c=1$ that \rf{*1.2}
is replaced by \cite{jain}
\beq{*3.x1}
\bar{n}_{N_T} (B) \sim \left[(N_T -B)B\right]^{\g-2}\left[\ln (N_T-B)
\ln B\right]^\a.
\eeq
In this formula we have left $\g$ and $\a$ as variables. Model calculations
give $\a=-2$, but it is not known whether this power is universal and
the model has not been solved analytically in the case of one gaussian field.

If we fit to \rf{*3.x1} in the way described above (including also
the $1/B$ correction) we extract for $c=1$
the following values of $\g$ and $\a$ for different lattice sizes:

\vspace{8pt}
\begin{center}
\begin{tabular}{ccc}
$N_{T}$  &  $\g$               & $\a$     \\ \hline
$1000$   &  $-0.22 \pm 0.05$ & $-0.5 \pm 0.4$    \\
$2000$   &  $-0.14 \pm 0.07$ & $-1.0 \pm 0.4$    \\
$4000$   &  $-0.09 \pm 0.08$ & $-1.2 \pm 0.4$
\end{tabular}
\end{center}
\vspace{8pt}

\noindent
Both $\g$ and $\a$  moves towards the expected values 0 and -2
as a function of $N_T$, but the finite size effects are clearly larger
here than for pure gravity.

In fig. 6 the results of a fit to \rf{*3.x1} for $c >1$ is included.
It is seen that $\g$ extracted in this way exceeds the theoretical
upper bound $\g=1/2$ (\cite{adfo}). In addition the power $\a$ decreases
from -1.2 for $c=1$ to -5 for $c=5$. We conclude that either logaritmic
corrections are not the right ones to include for $c >1$ or finite size
effects are so large that they make the fits unreliable.

What {\it is} clear from the analysis is that $\g$ increases with $c$
for $c$ in the range $0-5$. According to \rf{*1.2} this means that the
number of baby universes of a given size will increase, i.e. the fractal
structure will be more pronounced with increasing $c$. We have illustrated
this in fig. 7, which shows two ``typical'' surfaces
corresponding to $c=0$ and $c=5$.
It should be emphasized that the pictures are only intended to
visualize the {\it internal} structure,
i.e. the connectivity of the surfaces\footnote{The surfaces
are constructed in the following way: Given the connectivity matrix of the
triangulation we choose arbitrary coordinates for the vertices in $R^3$.
Next we introduce an attraction between neighbouring
vertices in order to keep the surface together and extrinsic curvature
to smooth out the surface during a Monte Carlo simulation.
When the surface reach a configuration without self-intersection
we put a pressure in the interior and
a Coulomb repulsion between distant vertices and in this way we
blow up the surface as a balloon.}

\newsection{Discussion}

We have verified that the technique of extracting the entropy exponent $\g$
directly from the distribution of baby universes is
superior to the methods used until now from a practical
point of view. A single (although long) Monte Carlo run for a fixed
value of $N_T$ is sufficient for extracting $\g$ and we get the
correct results for $c<1$. On the other hand the situation in the case
$c >1$ has not really improved much compared to the earlier measurements
\cite{abk}. We get in fact similar results, and this shows that
the method also works in the case $c>1$ and the ambiguity in extracting
$\g$ for $c >1$ is that we do not know the correct functional form
to be used in the fits. It is clear that it would be most interesting
if we could reverse the procedure and use the data to obtain knowledge
about the corrections to \rf{*1.1} for $c >1$. Our data are not yet good
enough to do this in a convincing way, but the problem is clearly not
due to the baby universe technique introduced in this paper, but due
to the inefficiency of the flip algorithm used to update the triangulations.

\vspace{24pt}

%\addtolength{\baselineskip}{-0.30\baselineskip}

\newpage

\begin{center}
{\large \bf Figure captions}
\end{center}

\begin{itemize}

\item[Fig.1] The distribution of baby universes in the case of
pure gravity.

\item[Fig.2] Fitted values of $\g_{B_0}$ for different cutoff's $B_{0}$ for
pure gravity.  Values are shown for fits with and without the
correction term included. The curve shows a fit using \rf{*ex1} resulting in
$\g = -0.496 \pm 0.005$ (errors are 95\% confidence limits for a
$\chi^2$-test).

\item[Fig.3] Fitted values of $\g$ vs the coupling $\b$ in the
case of one Ising model coupled to gravity.  Results are shown
for two lattice sizes, $N_{T} =1000$ and $2000$.

\item[Fig.4] Fitted values of $\g_{B_0}(\b)$ for various $\b$ as a function
of the cut-off $B_0$. The curves represent fits to \rf{*ex2}.

\item[Fig.5] The distributions of baby universes for up to five
Gaussian fields coupled to $2d$ gravity.  The values are normalized
with the distribution for pure gravity.

\item[Fig.6] Fitted values of $\g$ vs central charge in the case
of multble Gaussian fields.  Results are shown for fits without (circles) and
with (squares) a logarithmic correction term included.

\item[Fig.7] 3d illustration of the fractal structure of the surfaces
for $c=0$ (fig. 7a) and $c=5$ (fig. 7b). $N_T=200$ is used.

\end{itemize}

\end{document}